# Phase diagram of La$_{2-x}$Sr$_x$CuO$_4$ and stripes

Kirill V. Mitsen & Olga M. Ivanenko

*Lebedev Physical Institute RAS, 119991 Moscow, Russia*

**An elementary model of La$_{2-x}$Sr$_x$CuO$_4$ explaining the features of its superconducting phase diagram and the characteristics of spin textures without any assumption of stripe formation is proposed. The foundations of this model are the mechanism of negative-U center formation proposed earlier, the supposition of rigid localization of doped holes, and the concept of specific ordering of doped ions. It is shown that within the framework of the proposed model the features of "stripe" textures of La$_{2-x}$Sr$_x$CuO$_4$ reflect exclusively the geometrical relations existing in a square lattice and the competition between different types of doped hole ordering. The detailed agreement between the calculated and experimental phase diagrams is the decisive factor favoring the mechanism of high-T$_c$ superconductivity in which superconducting pairing arises due to scattering processes with intermediate virtual bound states of negative-U centers.**

La$_{2-x}$Sr$_x$CuO$_4$ is one of the simplest high-T$_c$ superconductors (HTSC), but its phase diagram shows a variety of different features still representing a puzzle for the theory. The interpretation of these features could clarify the nature of HTSC and mechanism of superconductivity in these compounds. Here we propose a model of La$_{2-x}$Sr$_x$CuO$_4$ that explains the details of the superconducting phase diagram and the characteristics of spin textures without any assumption of stripe formation.

It is known that HTSC reveal nonhomogeneous properties down to the nanoscopical scale. In our opinion this results from the strong localization of doped charges in the near vicinity of doped ions. This localization results in local variations of electronic structure of the parent charge-transfer insulator depending on the local mutual arrangement of doped charges.

Earlier we proposed[1,2] the mechanism of negative-U center (NUC) formation in HTSC. According to the model[1,2] the NUC are formed on pairs of neighboring Cu-ions in the CuO₂ plane. Figure 1a shows the electronic spectrum of undoped HTSC. Here the charge-transfer gap $\Delta_{ct}$ corresponds to the transition of electron from oxygen to nearest Cu ion. The originating hole is extended over 4 surrounding oxygen ions (Fig. 1b) due to overlapping of their $2p_{x,y}$ orbitals. This formation ($3d^{10}$-electron+2p-hole) resembles a hydrogen atom. We have shown[1,2] that the energy of two such excitations can be lowered (Fig. 1c) if such two side-by-side pseudo-atoms form a pseudo-molecule (Fig. 1d). This is possible due to formation of a bound state (of the Heitler-London type) of two electrons and two holes that emerge in the immediate vicinity of this pair of Cu ions.

It is follows from the experiment[3,4] that the doped hole is localized in the CuO₂ plane on four oxygen ions belonging to an oxygen octahedron adjacent to a Sr ion. That is, the holes in the central CuO₂ plane (Fig. 2a) are doped by Sr ions belonging to the LaO planes I and IV. We have distinguished[1,2] the different types of mutual arrangement of two doped holes, which determine the local properties of the crystal in the concentration region x<0.25. Simple estimation shows that the doped hole localized on oxygen ion is able to reduce the $\Delta_{ct}$ value for the four nearest Cu ions by ~1.6-1,8 eV. If two localized doped holes are spaced $3a$ or $a\sqrt{5}$, where $a$ is the lattice constant in CuO₂ plane (Fig. 2b,c), they reduce $\Delta_{ct}$ for the interior pair of nearest Cu ions to the extent that a two-electron transition is possible to this pair of Cu ions from neighboring oxygen ions while one-electron transitions are still impossible. These arrangements correspond to the formation of NUC on the interior pair of nearest Cu ions[1,2].

It is important that in the intermediate case, where the doped holes are spaced $a\sqrt{8}$ (Fig. 2d), no pairs of the nearest Cu ions can emerge, each having a nearby doped hole, i.e. an NUC is not formed. If doped holes are spaced $2a$ (Fig. 2e), then for the inner Cu ion the gap $\Delta_{ct}$ vanishes for single-electron transitions, too. This last case corresponds to that of ordinary metal.

## Ordering and percolation in La$_{2-x}$Sr$_x$CuO$_4$

As follows from the picture above-stated the HTSC on the way from insulator to metal passes through the particular doping range where local transfer of singlet electron pairs from oxygen ions to NUC are allowed while the single-electron transitions are still forbidden. NUC act as pair acceptors and generate additional holes, which are localized in the vicinity of the NUC. Conduction occurs in such a system if these regions of localization of hole pairs form a percolation cluster in the CuO$_2$ plane or, in other words, if the percolation threshold along NUC is exceeded. Given a random distribution of doped holes, the simultaneous coexistence of all types of doped hole arrangement becomes possible. Therefore it is difficult to expect the existence of large clusters whose properties are determined by one of a number of possible doped hole arrangements. However, as we wish to show, in La$_{2-x}$Sr$_x$CuO$_4$ the doped holes occupy the sites of the certain square lattices with lattice constants depending on $x$, which are sub-lattices of CuO$_2$ lattice. This results in the formation of different percolative broken lines with links whose length $l_{com}$ is commensurable with $a$. According to our consideration the broken lines with links $l_{com}$=3 or $\sqrt{5}$ form percolative clusters of NUC and correspond to the HTSC-phases, the cluster with links $l_{com}$=$\sqrt{8}$ is the insulator and the cluster with links $l_{com}$=2 is conventional metal. Lattices with $l_{com}$>3 correspond to the insulator.

In the considered scheme of the doping the system of Sr ion and hole located in oxygen sheet, represents an electric dipole that interacts with other dipoles through long range Coulomb potential. In such systems the orientation interaction between dipoles arises, that results in alignment of these dipoles by opposite poles to each other. Based upon crystal structure of $La_{2-x}Sr_xCuO_4$ it is possible to assume that replacement La on Sr will occur so that arising dipoles form vertical chains (reminding crankshafts) extended along $c$-axis (Fig. 3). Such arrangement simultaneously removes a question on what Sr ion (from planes I, IV or II, III on Figure 2a) dopes hole in the central $CuO_2$ plane.

Let believe that chains are flat and drawn up parallel to each other. The calculation of electrostatic interaction energy of dipole chains shows, that two nearest chains (Fig. 3a) will attract to each other if doped holes are spaced $l_{com} \geq 2$ and the next nearest chains repulse from each other. Such character of interaction results in dipole chain ordering and occupation of the sites in square lattice with certain parameter $l_{com}$ by doped holes.

It is follows from calculation that the energies of configurations with $l_{com}=2$, $\sqrt{5}$, $\sqrt{8}$ and 3 are close with a precision of $\sim 10^{-2}e^2/\varepsilon a$ per dipole ($\varepsilon$ - is a dielectric constant). Therefore the simultaneous coexistence of microdomains, in which doped holes occupy the sites in square lattices with different $l_{com}$, is possible.

Domains with given $l_{com}$ can exist only in the certain concentration range. This range is bounded above by the value $x_{com}=1/l_{com}$, above that the existence of physically significant domains with a given $l_{com}$ does not answer a condition of the mean concentration constancy.

At $x<x_{com}$ chains of dipoles are disrupted that results in appearance of vacancies in square lattices of doped holes. The microdomains with given $l_{com}$ will be kept up to some value $x=x_l$ which at random distribution corresponds to the threshold 2D vacancy percolation $c_v=1-x_l=0,593$[5]. Accordingly, the existence of domains with given $l_{com}$ is possible in the concentration range satisfying a condition

$$0,407/l^2_{com}<x\leq 1/l^2_{com}$$

The concentration of occupied sites $p$ in such microdomain changes with $x$ from $p\approx 0,4$ (at $x=0,407/l^2_{com}$) up to $p=1$ (at $x=1/l^2_{com}$).

The size of the ordered microdomain depends on the proximity of $x$ to $x_{com}$ and grows up to 200-600Å in *ab*-plane at x→0.12[6]. Along the *c*-axis the size of the ordered microdomain appears to be of some lattice constants, with the type of ordering of doped holes being repeated in every second $CuO_2$ plane.

At small $x$ (at mean distance between doped holes $l>3$) the chains of dipoles are believed to be in the planes parallel to both *c*- axis and orthorhombic *a*-axis so that the distance between holes along *a*-axis would be $\sqrt{8}$, i.e. would correspond to the minimum of energy of interaction.

**Phase diagram of $La_{2-x}Sr_xCuO_4$.** Let's find the regions of site percolation for microdomains with various $l_{com}$. Figure 4a shows the intervals of concentrations corresponding to percolation along the sites in microdomains with $l_{com}=3$, $\sqrt{8}$, $\sqrt{5}$ и 2 (in units of $CuO_2$ lattice constant *a*), i.e., intervals within which, according to above mentioned, there can exist percolative broken lines of doped holes with link lengths $l_{com}=3$ and $\sqrt{5}$ (NUC chains), $l_{com}=\sqrt{8}$ (insulator) and $l_{com}=2$ (conventional metal). Here, the solid rectangles depict the boundaries of the regions of percolative broken lines with a link length $l_{com}$. The heavy lines confine the areas of percolation over NUC with $l_{com}=3$ and $\sqrt{5}$.

Figure 4(b) shows the experimental phase diagram $T_c(x)$ for $La_{2-x}Sr_xCuO_4$ from [Ref. 7] The coincidence of the regions of superconductivity in the experimental phase diagram with intervals of percolation for $l_{com}=\sqrt{5}$ и 3 proves the conclusion that the fragments in question, which include pairs of neighboring Cu ions in $CuO_2$ plane together with two neighboring doped holes, are responsible for superconductivity in $La_{2-x}Sr_xCuO_4$. Note that the "dip" in $T_c(x)$ in the range $0.11<x<0.12$ (for $La_{2-x}Sr_xCuO_4$), caused by the absence of percolation over the chains of NUC is superimposed on the region of existence (as $x \to 1/8$) of $\sqrt{8} \times \sqrt{8}$ lattice of doped holes corresponding to the insulating phase. Further we show that just this feature makes it possible to observe a static incommensurable magnetic texture in this region.

**Stripes**

The incommensurable modulation[8,9] of AFM spin structure is observed in neutron scattering experiments[10] as two incommensurate peaks shifted in relation to AFM vector by $\varepsilon=1/T$ along the modulation vector. Here, $T$ is the period of the magnetic structure in units of the lattice constant.

The results of neutron studies[10-16] of the magnetic texture of $La_{2-x}Sr_xCuO_4$ can be summed up in the form of a stripe phase diagram (Fig.5a). It is seen that the incommensurate elastic-scattering peaks related to static modulation (shaded areas in Fig.5a) are observed at $x \leq 0.07$. In the range $0.07<x<0.15$ incommensurate peaks are observed in inelastic neutron scattering, which are evidence of dynamic modulation of spin texture (open areas in Fig.5a). At $x<0.07$ there are one-dimensional "diagonal" stripes with a single modulation vector directed along orthorhombic axis b, while at $x>0.05$ there is modulation in two directions parallel to the tetragonal axes ("parallel" stripes). To compare the spin structures for the cases of diagonal and parallel stripes, the diagonal stripes are examined in tetragonal units, too. Thus the spin modulation incommensurability parameter $\delta=\varepsilon$ for parallel and $\delta=\varepsilon/\sqrt{2}$ for diagonal stripes.

The theory, however, faces significant difficulties in describing the entire set of experimental results. The chief ones are:

1. the transition from diagonal to parallel stripes at $x\approx0.05$;

2. the relation $\delta\approx x$ for $x<0.12$ and $\delta\approx$const for $x>0.12$;

3. the change from static to dynamic stripes at $x\geq0.07$ and emergence of static correlations within a narrow region of concentrations at $x\approx0.12$ ("pinning of stripes");

4. the one-dimensionality of diagonal stripes and the two-dimensionality of parallel stripes;

5. the slant of parallel stripes.

In an attempt to explain the results of neutron experiment Gooding et al.[17,18] proposed the spin-glass model based on the supposition of chaotic distribution of localized doped holes. Supposedly a localized doped hole generates a long-range field of spin distortions of AFM background that can be described as the creation of a topological excitation, a skyrmion[19,20], with topological charge ±1 corresponding to twisting of AFM order parameter in the vicinity of a localized hole. Thus, doping destroys long range AFM order and leads to the formation of AFM-ordered microdomains whose angular points are specified by doped holes.

Combining some ideas from[17,18] together with our ideas concerning the mechanism of NUC formation and Sr ordering we provide below an alternative explanation of the observed spin and charge modulations.

**Parallel stripes.** Let us examine the case of complete ordering at $x_{com}$=1/8. We assume that each hole circulates over the oxygen sheet surrounding a copper ion and that because of the interaction between the hole current and the spins of the four nearest copper ions the latter are polarized.

Figure 6a shows a possible ordering of the projections of Cu spins on the $CuO_2$ plane for a completely ordered arrangement of localized holes at $x$=1/8. Here, the $CuO_2$ plane is broken up into separate AFM-ordered quadrangular microdomains whose corners are determined by the localized doped holes. The directions of Cu spin projections at lattice sites are indicated by arrows. The emerging pattern is characterized by the AFM-ordering of the microdomains proper and by the ordered alternation of skyrmions. Such a checkerboard pattern of AFM-ordered microdomains (Fig. 6a) results in an imitation of a "parallel stripe" structure. Actually, the "parallel stripes" are of the chains antiphase microdomains with magnetization directions along horizontal or vertical axes (Fig. 6a). The magnetic modulation period T=8 in this case is equal to the total size of two antiphase microdomains along the modulation vector and a rectangle with an area equal to *2×T* must contain two sites. I.e.

*2Tx*=2; and δ=1/T=*x*=1/8

Thus, the relation *δ=x* in the case of parallel stripes is caused exclusively by the fact that doped holes lie along straight lines equally spaced at *l*=2.

Now let us consider the magnetic textures formed at a slight deviation of *x* below $x_{com}$=1/8. Kimura et al.[16] used an $La_{1.88}Sr_{0.12}CuO_4$ sample to observe the modulation of a spin texture with an incommensurability parameter $\delta = 0.118$. This corresponds to a mean texture period $T \approx 8.5$ (in units of a), i.e., to the alternation of two periods, $T_1=8$ and $T_2=9$. Figure 6b shows the picture of an ordered distribution of doped holes we proposed for a mean concentration $x=0.118$, which was obtained by cutting the completely ordered lattice at $x=1/8$ along the orthorhombic $a$-axis and shifting one part in relation to the other by the vector q=(1,1). Such a dislocation conserves the coherence of ordering in domains on both sides of the dislocation (with a small phase shift). Such a structure (Fig. 6b) produces characteristic reflections in the diffraction pattern, and these reflections correspond to incommensurable modulation of both spin with an incommensurability parameter $\delta$. The condition of conservation of the mean concentration yields

$$T_d x_m = (T_d - 1)x_l$$

Here $T_d$ is the mean dislocation period in units of $a$. To maintain the mean concentration $x_m=0.118$ for a local concentration inside a domain $x_l=0.125$, the introduced diagonal dislocations must have a mean period $T_d =17$ ($=T_1+T_2$). Such quasiperiodic dislocations, which lead to incommensurable modulation of the crystal structure and the spin texture, result in incommensurable reflections $\delta = 0.118$ in full agreement with the experiment[16].

The pattern of ordering is so special in that, as Fig. 6b shows, the parallel stripes are shifted by one lattice constant, i.e., as though, they deflected from tetragonal axes by an angle $\theta_\gamma =1/17 \approx 3.3°$ toward the orthorhombic axis $\boldsymbol{b}$. These are those "slanted" parallel stripes with a slope angle of 3°, which were observed by Kimura et al.[16].

Let us now turn to the case of arbitrary values of $x$ for $x<0.125$. Here, the distribution pattern can be obtained from the completely ordered distribution at $x=0.125$ (Fig. 6a) by removing a certain number of sites one after another. The texture imitating parallel stripes may occur down to x≈0.05 because small clusters (pieces of broken lines) with $l_{com}=\sqrt{8}$ still survive.

Let us suppose that the lattice contains such correlated remnant fragments of a parallel stripe texture genetically linked to $\sqrt{8}\times\sqrt{8}$ microdomains. The related neutron diffraction pattern exhibits characteristic reflections determined by the mean remnant texture period. In turn, the mean period $T$ of this texture, defined as the distance between the equivalent points of unidirectional magnetic domain, includes two occupied sites. I.e. a rectangle with an area equal to $T \cdot l_{com}/\sqrt{2} =2T$ must contain two sites and $\delta=1/T=x$. This relation for parallel stripes takes place at 0.05<x<0.125, when the $\sqrt{8}\times\sqrt{8}$ lattice is occupied.

**Diagonal stripes.** As is seen from Figure 6b the inserted dislocations are actually the nucleus of diagonal stripes extended along *a*-axis. They became apparent as quasi-periodic structures at x<0,05 when the remains of $\sqrt{8}\times\sqrt{8}$ microdomains disappear; but diagonal chains of doped holes equally spaced at $l_{com}=\sqrt{8}$ are left intact. Therefore diagonal stripes will be directed always along the orthorombic *a*-axes with modulation vector, accordingly, along the orthorombic *b*-axes. If all dipoles are ordered in diagonal ranks, the period of diagonal spin modulation $T$ (in tetragonal axes) should be equal $T=1/\sqrt{2}\,x$ (or $\delta=\sqrt{2}\,x$). Since the part of doped holes can remain in space between dipole ranks the period of observable spin structure will be more, than $1/\sqrt{2}\,x$, accordingly $\delta$ is less than $\sqrt{2}$ x. The experimental value of $\delta$ varies[11] from $\delta\approx0.7x$ up to $\delta\approx1.4x$ over the range 0.01<x<0.05.

**Dynamic stripes.** The last problem that we will discuss deals with dynamic stripes. Figure 5b shows the concentration ranges within which there can be antiferromagnetically correlated clusters of $\sqrt{8}\times\sqrt{8}$ microdomains (0.05<x<0.125) and diagonal chains of doped holes spaced $l=\sqrt{8}$ (*x*<0.07). The dashed lines limit the regions of existence of percolation clusters with $l_{com}=3$ and $l_{com}=\sqrt{5}$. Such conductive cluster bordering on AFM cluster destroys the static spin correlations because of the motion of charges that disrupt the magnetic order in the neighborhood along its path. Therefore spin correlations can be observed in the range $0.66<x<0.11$ only in inelastic neutron scattering as dynamic incommensurable magnetic fluctuations. What is remarkable (see Fig. 5b) is that in addition to the region $x<0.07$ there is a narrow interval of concentrations $0.11<x<0.12$ where there is no percolation along NUC, and it is precisely in this interval that static incommensurable correlations are again observed (Fig. 5a). At nonhomogeneous Sr distribution the existence of small ordered domains with $x\leq0.125$, bordering a percolative cluster with $l_{com}=\sqrt{8}$, is possible at $x>0.125$. It results in the experimental observation of dynamic spin texture with $\delta\approx0.125$.

**Conclusion.**

We have presented a model of $La_{2-x}Sr_xCuO_4$, which makes it possible to provide a detailed explanation of the superconducting and magnetic phase diagrams of this compound without an assumption of electronic phase separation or stripe formation. It was shown that the features of the phase diagrams reflect only the geometrical relations existing in a square lattice and the competition of different types of Sr ordering. The close agreement between the calculated phase diagrams and the experimental results may serve as an important argument in favor of the model of high-$T_c$ superconductors proposed in [Ref. 1] and it is the decisive factor favoring the mechanism of high-$T_c$ superconductivity in which superconducting pairing arises because of renormalization of electron-electron interaction due to scattering processes with intermediate virtual bound states of NUC[21-25].

**References**

1.  Mitsen, K.V. & Ivanenko, O.M. Phase diagram of $La_{2-x}M_xCuO_4$ as the key to understanding the nature of high- $T_c$ superconductors. *Uspekhi Fizicheskikh nauk* **174**, 545-563 (2004) [*Physics-Uspekhi* **47**, 493-510 (2004)].

2.  Mitsen, K.V. & Ivanenko, O.M. Phase diagram features and "stripe" structure characteristics of $La_{2-x}Sr_xCuO_4$ as consequences of specific dopant ordering. *Physica C* **417**, 157-165 (2005).

3.  Haskel, D., Polinger, V. & Stern, E.A. Where do the doped holes go in LaSrCuO? A close look byXAFS. *AIP Conf. Proc.* **483**, 241-246 (1999).

4.  Hammel, P.C., Statt, B.W., Martin, R.L., Chou, F. C., Johnston, D. C. & Cheong, S-W. Localized holes in superconducting lanthanum cuprate. *Phys. Rev.* B **57**, R712-R715 (1998).

5.  Ziff, R.M. Spanning probability in 2D percolation. *Phys. Rev. Lett.* **69**, 2670-2673 (1992).

6.  Savici, A.T., Fudamoto, Y., Gat, I.M., Ito, T., Larkin, M.I., Uemura, Y.J., Luke, G.M., Kojima, K.M., Lee, Y.S., Kastner, M.A., Birgeneau, R.J. & Yamada, K. Muon spin relaxation studies of incommensurate magnetism and superconductivity in stage-4 $La_2CuO_{4.11}$ and $La_{1.88}Sr_{0.12}CuO_4$. *Phys. Rev. B* **66**, 014524-014537 (2002)

7.  Kumagai, K., Kawano, K., Watanabe, I., Nishiyama, K. & Nagamine, K. Magnetic order and evolution of the electronic state around x=0.12 in $La_{2-x}Ba_xCuO_4$ and $La_{2-x}Sr_xCuO_4$. *J. Supercond.* **7**, 63-67 (1994).

8.  Zaanen, J. & Gunnarsson, O. Charged magnetic domain lines and the magnetism of high-Tc oxides. *Phys. Rev. B* **40,** 7391-7394, (1989).

9. Kivelson, S.A., Emery, V.J. & Lin, H-Q. Doped antiferromagnets in the weak-hopping limit. *Phys. Rev.* B **42,** 6523-6530 (1990).

10. Tranquada, J.M., Axe, J.D., Ichikawa N. Y. Nakamura, Y., Uchida, S. & Nachumi B. Neutron-scattering study of stripe-phase order of holes and spins in $La_{1.48}Nd_{0.4}Sr_{0.12}CuO_4$ *Phys. Rev. B* **54,** 7489-7499 (1996).

11. Yamada, K., Lee, C.H., Endoh, Y., Shirane, G., Birgeneau, R.J. & Kastner M.A. Low-frequency spin fluctuations in the superconducting $La_{2-x}Sr_xCuO_4$. *Physica C* **282-287**, 85-88 (1997).

12. Yamada, K., Lee, C.H., Kurahashi, K., Wada, J., Wakimoto, S., Ueki, S., Kimura, H., Endoh, Y., Hosoya, S., Shirane, G., Birgeneau, R.J., Greven, M., Kastner, M.A. & Kim, Y.J. Doping dependence of the spatially modulated dynamical spin correlations and the superconducting-transition temperature in $La_{2-x}Sr_xCuO_4$. *Phys. Rev. B* **57**, 6165-6172 (1998).

13. Matsuda, M., Fujita, M., Yamada, K., Birgeneau, R.J., Endoh, Y. & Shirane, G. Electronic phase separation in lightly doped $La_{2-x}Sr_xCuO_4$. *Phys. Rev. B* **65**, 134515-134520 (2002).

14. Fujita M., Yamada K., Hiraka H., Gehring P.M., Lee S.H., Wakimoto S. & Shirane G. *Phys. Rev. B* **65**, 64505-64511 (2002).

15. Fujita, M., Goka, H., Yamada, K. & Matsuda, M. Structural effect on the static spin and charge correlations in $La_{1.875}Ba_{0.125-x}Sr_xCuO_4$. *Phys. Rev. B* **66**, 184503-184508 (2002).

16. Kimura, H., Matsushita, H., Hirota, K., Endoh, Y., Yamada, K., Shirane, G., Lee, Y.S., Kastner, M.A. & Birgeneau, R.J. Incommensurate geometry of the elastic magnetic peaks in superconducting $La_{1.88}Sr_{0.12}CuO_4$. *Phys. Rev. B*. **61**, 14366-14369 (2000).

17. Gooding, R.J., Salem, N.M. & Mailhot, A. Theory of coexisting transverse-spin freezing 2 and long-ranged antiferromagnetic order in lightly doped $La_{2-x}Sr_xCuO_4$. *Phys. Rev. B* **49,** 6067- 6073 (1994).

18. Gooding, R.J., Salem, N.M., Birgeneau, R.J. & Chou, F.C. Sr impurity effects on the magnetic correlations of $La_{2-x}Sr_xCuO_4$. *Phys. Rev. B* **55**, 6360-6371 (1997).

19. Belavin, A.A. & Polyakov, A.M. Metastable states of two-dimentional isotropic ferromagnets. *Pis'ma Zh. Eksp. Teor. Fiz.* **22**, 503-506 (1975) [*JETP Lett.*, **22**, 245-248 (1975)].

20. Gooding, R.J. Skyrmion ground states in the presence of localizing potentials in weakly doped $CuO_2$ planes. *Phys. Rev. Lett.* **66**, 2266-2269 (1991).

21. Simanek, E. *Solid State Commun*. Superconductivity at disordered interfaces. **32,** 731-734 (1979).

22. Ting, C.S., Talwar, D.N. & Ngai, K.L. Possible Mechanism of Superconductivity in Metal-Semiconductor Eutectic Alloys. *Phys.Rev.Lett.* **45,** 1213-1216 (1980).

23. Schuttler, H.-B., Jarrell, M. & Scalapino, D.J. Superconducting $T_c$ enhancement due to excitonic negative-$U$ centers: A Monte Carlo study. *Phys.Rev.Lett.* **58,** 1147-1149 (1987).

24. Ranninger, J. & Robin, J.M. The boson-fermion model of high-$T_c$ superconductivity. Doping dependence. *Physica C* **253,** 279-291 (1995).

25. Arseev, P.I. Possible mechanism for high temperature superconductivity. *Zh. Eksp. Teor. Fiz.* **101,** 1246-1258 (1992) [*Sov. Phys. JETP* **74,** 667-680 (1992)].Style tag for references (e.g. 1. Stacey, F. D., Brown, K. & Smith, L. Maintenance of genomic methylation. *Mol. Cell Biol.* **32,** 123–125 (1999).).

**Acknowledgements**. We thank V.L. Ginsburg , E.G. Maksimov and A.A. Alexandrov for stimulating discussions. This work was supported by grant from Russianт Federal Agency of Science and Innovations.

Correspondence and requests for materials should be addressed to K.V. Mitsen (e-mail: mitsen@sci.lebedev.ru).

**Figure captions:**

Fig. 1. The mechanism of negative-U center formation. (a) Electron spectrum of undoped cuprate HTSC. $U_H$ is the repulsive energy for two electrons on Cu ion, $t_{oo}$ is the overlapping integral between $2p_{x,y}$ states of neighboring oxygen ions. The charge-transfer gap $\Delta_{ct}$ corresponds to the transition of electron from oxygen to nearest Cu ion with the origin of hole extended over 4 surrounding oxygen ions (Fig. 1b). (c) The energy of two such excitations can be lowered if two side-by-side "hydrogen" pseudo-atoms form "hydrogen" pseudo-molecule (Fig. 1d). $\Delta E_U$ is the binding energy.

Fig. 2. The types of arrangement of two nearest localized doped holes in $CuO_2$ plane. (a) The unit cell of $La_{2-x}Sr_xCuO_4$; (b,c) if the doped holes are spaced $3a$ or $a\sqrt{5}$ the NUC is formed on the interior pair of nearest Cu ions; (d) if the doped holes are at a distance $l=a\sqrt{8}$, no pairs of the nearest Cu ions can emerge each having the doped hole in the neighborhood, i.e. an NUC is not formed; (e) if doped holes are spaced $2a$, then for the inner Cu ion the gap $\Delta_{ct}$ vanishes for one-electron transitions, too. This case corresponds to the ordinary metal.

Fig. 3. The picture of Sr ions ordering: negatively charged Sr ions together with "attributed" to them doped holes are the dipoles which draw up by the opposite poles to each other forming the chains of dipoles reminding crankshafts.

Fig. 4. Superconducting phase diagram of $La_{2-x}Sr_xCuO_4$. a) The concentration intervals corresponding to percolation along the sites in domains with different $l_{com}$. Solid rectangles depict the boundaries of the regions of percolatiive broken lines with a link length $l_{com}$=3, $\sqrt{8}$, $\sqrt{5}$ и 2. The heavy lines confine the areas of percolation over NUC with $l_{com}$=3 and $\sqrt{5}$. b) The experimental phase diagram $T_c(x)$ for La$_{2-x}$Sr$_x$CuO$_4$ [Ref.7].

Fig. 5. Magnetic phase diagram of La$_{2-x}$Sr$_x$CuO$_4$. (a) Experimental "stripe phase" diagram of La$_{2-x}$Sr$_x$CuO$_4$ [Ref.10-16]. The angle $\alpha$=45$^o$ corresponds to the concentration range where diagonal stripes are observed, while $\alpha$=90$^o$ corresponds to the concentration range within which vertical stripes are observed. The hatched regions correspond to the intervals where static stripes were observed. (b) The dashed lines confine the regions of percolation over NUC with $l$=3 and $l$=$a\sqrt{5}$ (dynamic stripes); the heavy lines confine the regions where $\sqrt{8}\times\sqrt{8}$ (0.05<x<0.125) microdomains and diagonal chains of doped holes spaced $l$=$\sqrt{8}$ (x<0.07) are existed. The numbers at the vertices of rectangles indicate the values of $l_{com}$ for a given percolation region.

Fig. 6. The imitation of stripe textures when doped hole ordering takes place. (a) Projections of spin for $x$= 1/8 when doped holes become ordered in $\sqrt{8}\times\sqrt{8}$ lattice. The chains of large shaded squares represent of the chains of antiphase microdomains with magnetization direction along horizontal axis – "horizontal stripes"; the chains of large white squares form "vertical stripes"; (b) The same as in (a) but for $x$<1/8. The plane is partitioned into two domains separated by diagonal dislocations. The shift of stripes by one cell, that appear on each dislocation, lead to an effective tilt of stripes with an angle $\theta\gamma$. Bold rectangles mark out the "vertical" and "horizontal stripes".

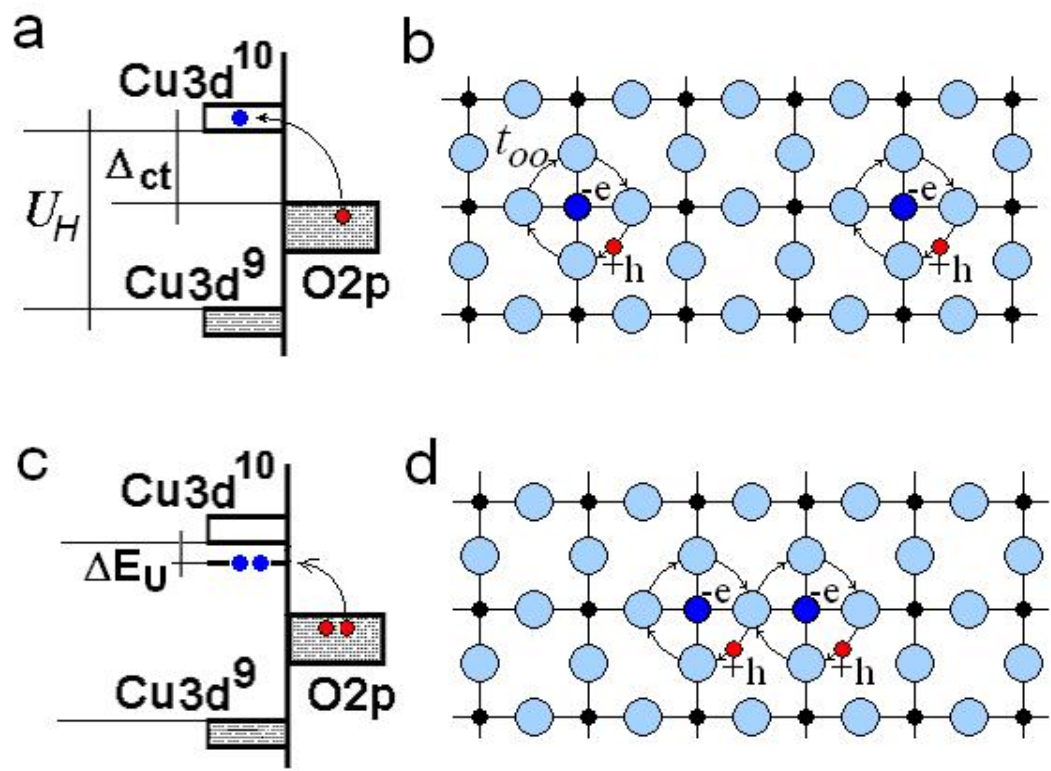

Fig. 1

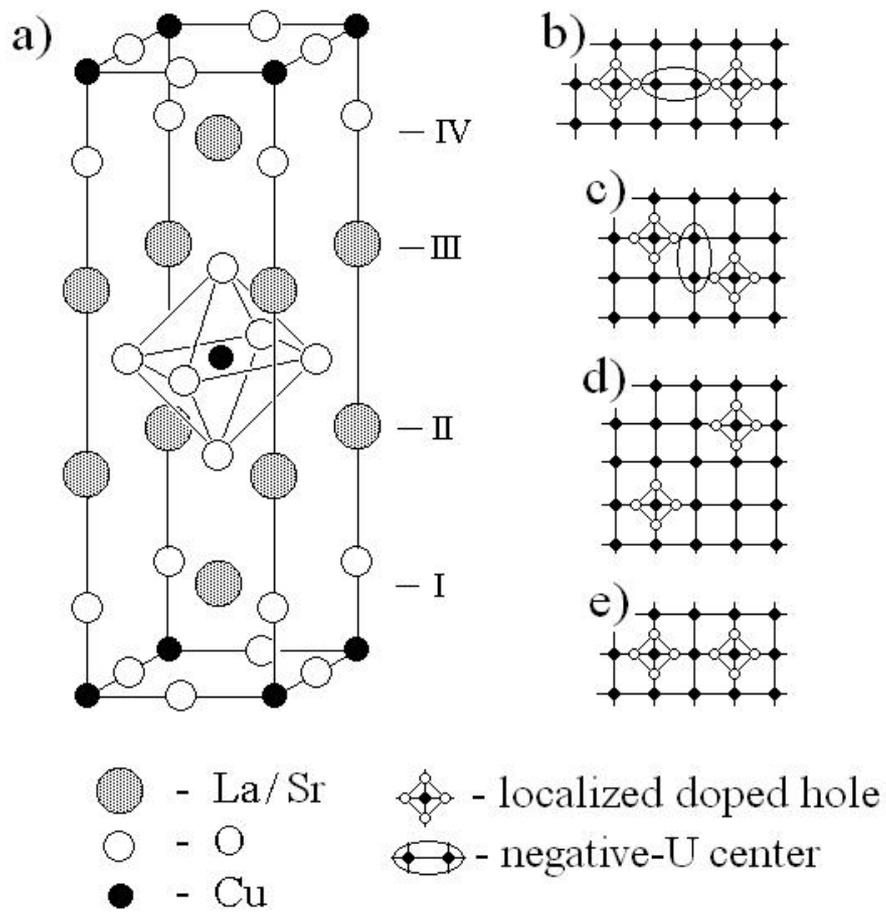

- La/Sr
- O
- Cu

⬦ - localized doped hole

⬯ - negative-U center

Fig. 2

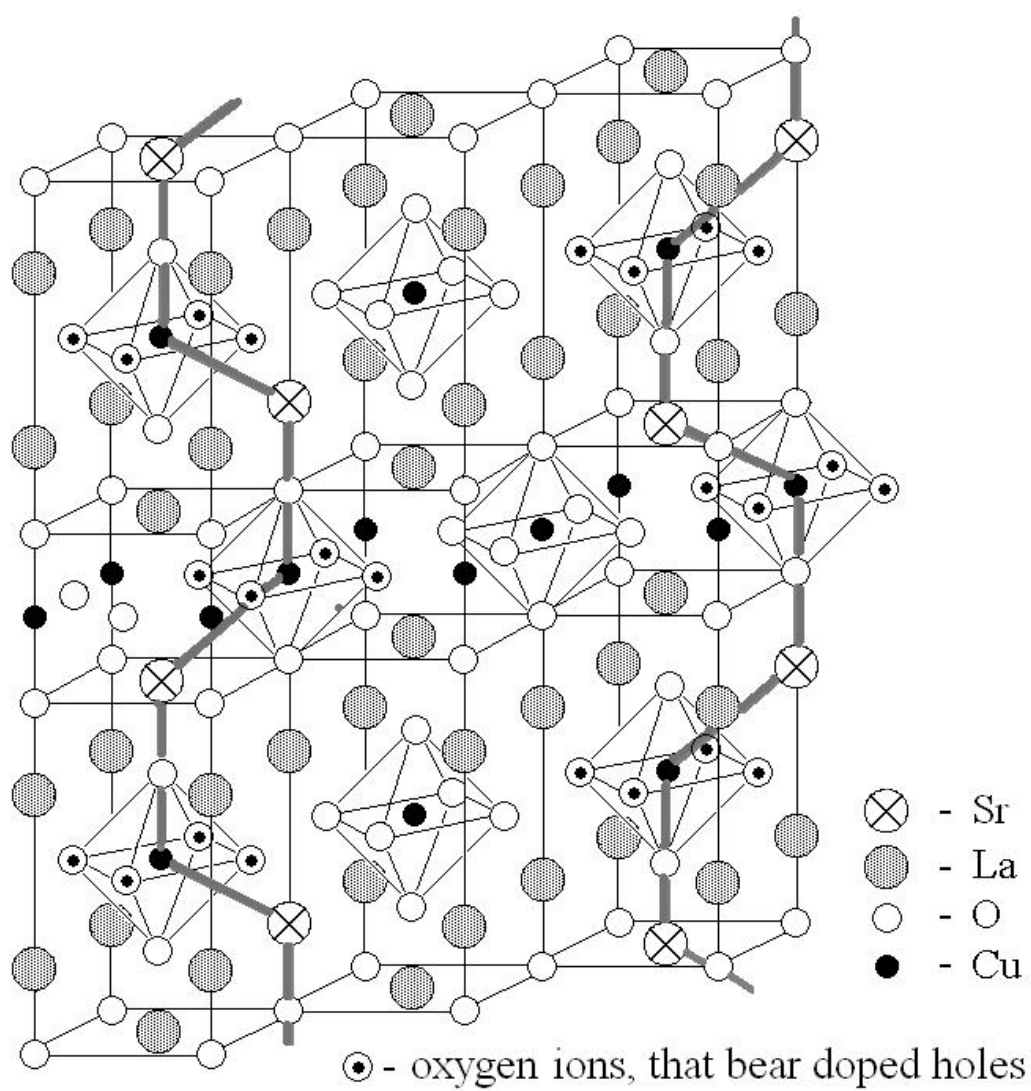

⊙ - oxygen ions, that bear doped holes

- Sr
- La
- O
- Cu

Fig. 3

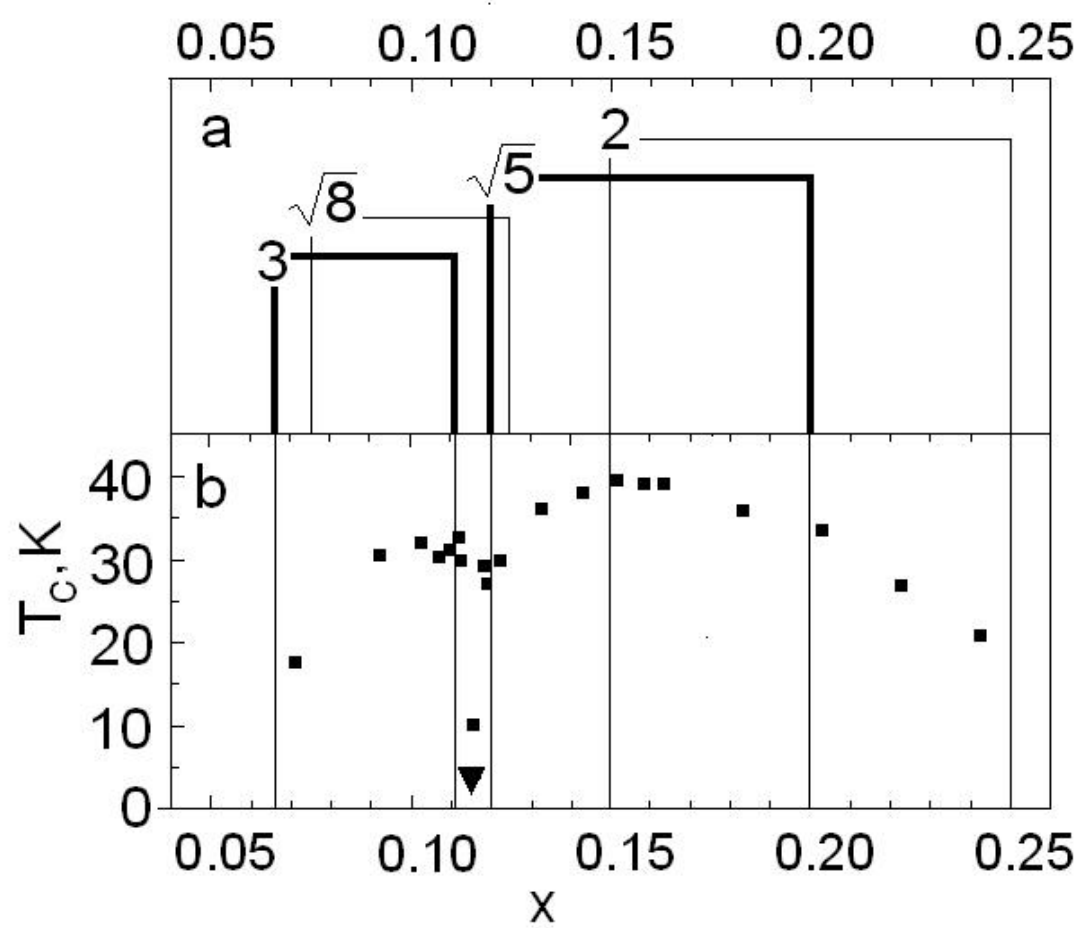

Fig. 4

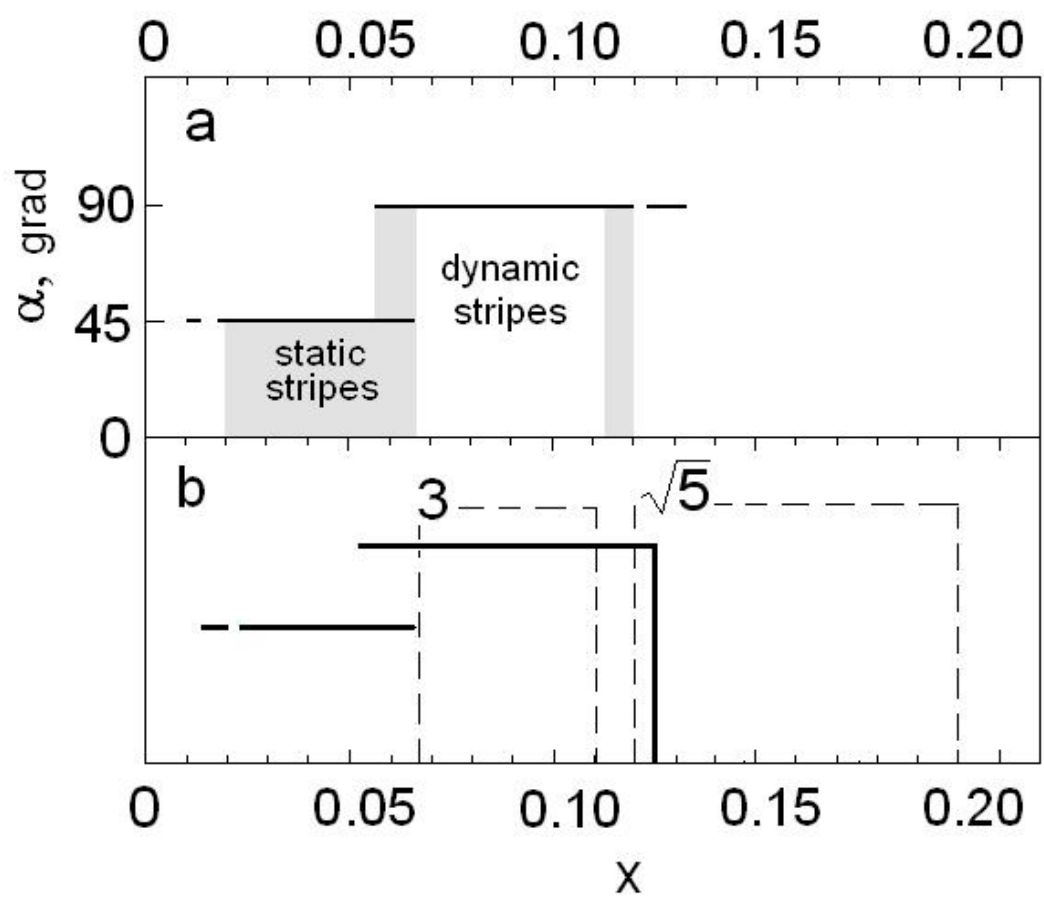

Fig. 5

Fig. 6